# ANALYZING PEER SELECTION POLICIES FOR BITTORRENT MULTIMEDIA ON-DEMAND STREAMING SYSTEMS IN INTERNET

Carlo Kleber da Silva Rodrigues[1,2]

[1]Electrical and Electronics Department, Armed Forces University – ESPE, Sangolquí, Ecuador
[2]University Center of Brasília – UniCEUB, Brasília, DF, Brazil

## ABSTRACT

*The adaptation of the BitTorrent protocol to multimedia on-demand streaming systems essentially lies on the modification of its two core algorithms, namely the piece and the peer selection policies, respectively. Much more attention has though been given to the piece selection policy. Within this context, this article proposes three novel peer selection policies for the design of BitTorrent-like protocols targeted at that type of systems: Select Balanced Neighbour Policy (SBNP), Select Regular Neighbour Policy (SRNP), and Select Optimistic Neighbour Policy (SONP). These proposals are validated through a competitive analysis based on simulations which encompass a variety of multimedia scenarios, defined in function of important characterization parameters such as content type, content size, and client´s interactivity profile. Service time, number of clients served and efficiency retrieving coefficient are the performance metrics assessed in the analysis. The final results mainly show that the novel proposals constitute scalable solutions that may be considered for real project designs. Lastly, future work is included in the conclusion of this paper.*

## 1. INTRODUCTION

The BitTorrent protocol is undoubtedly an effective peer-to-peer (P2P) solution for file (content) replication over Internet [1, 2, 3, 4]. The overall idea of this protocol is that peers must cooperate with each other to replicate the wanted file. In case a peer does not contribute at all, it will not be able to download the file. This protocol is simple, efficient, and robust [3, 4]. This is why it has been widely used for various purposes, such as for distributing large software packages [5, 6, 7].

The files that are distributed using BitTorrent are first split into pieces and each piece is then split into blocks. The piece selection and peer selection policies, respectively, are the basis of the BitTorrent protocol. The first policy is employed by a peer to decide which file pieces are going to be requested for download. The second policy, by its turn, is employed by a peer to choose the peers that will receive the pieces it owns [8, 9].

The widespread popularity of BitTorrent has motivated a lot of proposals to adapt it in order to meet the requirements of Internet on-demand streaming systems. Most of these proposals are based on the modification of both the original piece and peer selection policies, respectively. However, the truth is that the piece selection policy has received much more attention than the peer selection policy [1, 5, 6]. Additionally, interactive scenarios have become tremendously popular on Internet. This means that, for most of the nowadays on-demand streaming applications, the client may behave as though he were at his home in front of a DVD-player, i.e., he is able to execute DVD-like actions during the multimedia file playback [1, 2, 5, 6].

The context discussed above gives the motivation for this article, whose main objective is to propose three novel peer selection policies for the design of BitTorrent-like protocols. These policies are denoted as Select Balanced Neighbour Policy (SBNP), Select Regular Neighbour Policy (SRNP), and Select Optimistic Neighbour Policy (SONP). They are destined for interactive on-demand streaming systems and are validated through simulations on a variety of multimedia file replication scenarios. These scenarios are categorized in function of content type, content size, and client´s interactivity profile. Three important performance metrics are computed in the analysis: *service time*, *number of clients served*, and *efficiency retrieving coefficient*. The final results show that the novel policies constitute efficient scalable solutions which may be used for real project designs.

The remainder of this text is organized as explained in the following. Section 2 briefly reviews several concepts related to the BitTorrent protocol and debates on the operation of its core algorithms: piece and peer selection policies, respectively. Besides, there is a general discussion of the reciprocity principle, which plays a very important role when adapting BitTorrent's peer selection policy to on-demand streaming scenarios. Section 3 presents related work. The novel proposals come in Section 4. The performance evaluation lies in Section 5, including the simulation setup, the competitive metrics used in the simulations, and all results and the corresponding analysis. At last, conclusions and future work appear in Section 6.

## 2. BASIS

### 2.1. Concepts and general overview

Understanding the BitTorrent protocol involves the notion of *torrent* and *swarm*. Depending on context, a torrent may refer either to a metadata file or to a set of contents, described by metadata files. In case it refers to a metadata file, it has the extension *.torrent* and contains metadata related to the content it makes downloadable from Internet. This metadata includes, e.g., name, length, hashing information, and the IP address of the so-called *tracker* [4, 8, 9].

The tracker is a central entity which coordinates the communication within the set of peers that cooperate to replicate a content of a torrent (i.e., set of contents) among each other. This set of peers is denoted as a swarm and, hence, is associated to a session of content transfer. Swarms are independent. In particular, participating in a swarm does not bring any benefit for the participation in another swarm. Peers involved in a swarm cooperate among each other using the so-called *swarming* techniques [4, 8, 9].

As previously mentioned, in order to be replicated, files are first split into pieces, and each piece is split into blocks. Pieces are typically 256 kB in size, while blocks are typically 16 kB in size [4, 8, 9]. To request data, the peer identifies the piece it wants to download and then requests a number of blocks (typically five) of this piece. Whenever a block arrives, a new block request is sent in such a way that there are always several requests (typically five) pipelined at once. This surely helps to avoid a delay between pieces being sent [4, 8, 9].

To say that a peer *A* is *interested* in another peer *B* means that this peer *B* has pieces that the peer *A* does not have [9]. To say that a peer *A* is *not interested* in a peer *B* means that this peer *B* has just a subset of the pieces of the peer *A* [9]. To say that a peer *A* *chokes* a peer *B* means that the peer *A* decides not to send data to the peer *B* [9]. Finally, to say that a peer *A* *unchokes* a peer *B* means that the peer *A* decides to send data to peer *B* [8, 9].

The *peer set* or *neighbour set* is a list of the other peers that a given peer knows about. This way, a swarm may be viewed as a collection of interconnected peer sets, each with typically a minimum of 20 and a maximum of 80 peers. It is worth saying that a peer is allowed to have only 40 initiated connections within the possible maximum 80 at each time. That is, the 40 remaining connections may be only initiated by its neighbours. However, a peer can only send data to a subset of its peer set, denoted as *active peer set*. The active peer set has a typical size of four peers. This overall policy guarantees a good interconnection balance among the peer sets in the torrent [4, 8, 9].

A peer may be in two states: the *leecher state*, when it is downloading content but does not own all the pieces yet; and the *seed state*, when the peer owns all the pieces of the content. Each peer is aware of the pieces of all the other peers belonging to its peer set. A *local peer* is the peer that wants to download pieces, while *remote peers* are the peers that are in the peer set of the local peer. The exchange of pieces is governed by two core algorithms: the *rarest first* and the *choke* algorithms, respectively [4, 8, 9].

The *choke algorithm* is the peer selection policy used in BitTorrent. The active peer set is determined by this algorithm, which is also denoted as *tit-for-tat algorithm*. Only peers that are interested in the local peer and unchoked by the local peer are part of the active peer set. In general, the peers that provide data at the highest speeds are favoured. The main goal of this algorithm is to guarantee a reasonable level of upload and download reciprocation [4, 8, 9]. A more detailed description of the peer selection policy of BitTorrent is given in the next section.

The *rarest first algorithm* refers to the normal operation of the piece selection policy used in BitTorrent. It is also called the *local rarest first algorithm*. Each peer has a *rarest pieces set*. The *rarest pieces* are the pieces that have the least number of copies in the peer set. In general, right after being unchoked, the local peer randomly requests the next piece to download in its rarest pieces set and considering the available pieces on the remote peer that unchokes it [8, 9]. A more detailed description of the piece selection policy is presented in the next section.

## 2.2. Operation of the BitTorrent

To participate in a swarm, a new peer, also denoted as a newcomer, first has to download the metadata file of extension *.torrent* from an ordinary web server. This peer then contacts the tracker and receives a random list *L* of other peers (both seeds and leechers) that already belong to the swarm. The list *L* usually has 50 peers randomly selected [8, 9].

To prevent excessive traffic at the tracker, the peer´s request rate is limited. The default value in the original BitTorrent tracker implementation from Cohen [8] allows a single request every five minutes. However, in Internet, various tracker implementations exist and, hence, much shorter time intervals between requests (e.g., 10 seconds) are likely to be allowed [3, 4].

After receiving the list *L*, the new peer then attempts to establish bidirectional persistent connections (TCP connections) with the peers from it. Those connections resulting successful form the peer set or neighbour set. Once the connections are established, the peers exchange information among themselves. For example, they communicate the *bitfield*, i.e., a map of the data pieces they may share. This initial peer set is augmented by peers connecting directly to the new peer [8, 9].

At this moment, the new peer gets to know the neighbours (i.e., remote peers) that are interested in it as well as the neighbours that it is interested in. To receive pieces, the new peer has first to be unchoked by neighbours it is interested in. Conversely, to send pieces, the new peer has first to unchoke neighbours that are interested in it. These two processes are ruled by the peer selection policy, explained in Section 2.2.1. Finally, to end this section, the sequence of events just discussed above is shown in Figure 1.

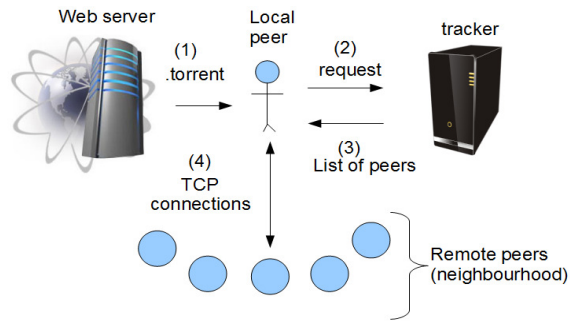

Figure 1. Sequence of events for a peer to join a torrent.

### 2.2.1. Peer selection policy

In this section, the peer selection policy of BitTorrent is explained. As already mentioned, this policy is based on the choke algorithm and determines how a peer selects another peer to upload data to. This is done by means of Algorithm 1, which is a step-by-step description of how this policy operates. It is explained from the point of view of the local peer that has just been told of its neighbours. In this description, $\delta$ is typically set to 10, *interested* means interested in the local peer, and *choked* means choked by the local peer. Lastly, since this policy implicitly has the role of incentivizing peer cooperation, it is usually designed to favour good uploaders.

In addition to the description given in Algorithm 1, it is worth saying that if the local peer is in leecher state and happens to go over $\tau$ seconds ($\tau$ is typically set to 60) without getting a single piece from a given remote peer from whom it was formerly downloading, the local peer assumes it is *snubbed* by that remote peer and does not upload to it anymore (i.e., the remote peer is set choked), except under an optimistic unchoke [8, 9].

In case the local peer is snubbed by all its remote peers from whom it was formerly downloading, it may then optimistically unchoke more than one remote peer [8, 9]. This is an exception to the exactly one optimistic-unchoke rule, mentioned in Step 4 of Algorithm 1. This will hopefully cause download rates to recover much more quickly when they falter [8, 9]. The overall understanding of this algorithm is depicted in Figure 2.

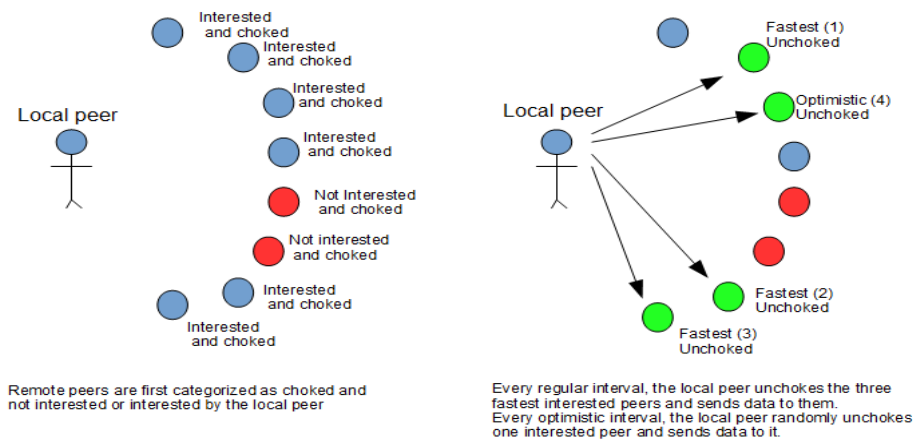

Figure 2. Operation of the peer selection policy of BitTorrent.

**Algorithm 1: Peer Selection Policy of BitTorrent**

Begin

Initialization

    Step 1 The neighbours of a local peer are all choked and categorized in either *interested* or *not interested* remote peers.

    Step 2 The local peer's upload capacity is divided into four upload slots: three slots are denoted as *regular unchoke slots*, and one slot is denoted as *optimistic unchoke slot*.

Repeat

    Step 3 Every $\delta$ seconds (this is denoted as the *Regular unchoke interval*)

        Step 3.1 If the local peer is in *leecher state*: all interested remote peers are ordered with respect to their upload rate to the local peer, and the three fastest peers are unchoked. In other words, the three regular unchoke slots are assigned, one for each of the three fastest peers [8, 9].

        Step 3.2 Else if the local peer is in *seed state*: all interested remote peers are ordered according to the upload rate the local peer has to them and the three fastest peers are unchoked [8, 9]. Note that, under this state, peers with high download rates are favoured independently of their contribution to the torrent.

    Step 4 Every $3\delta$ seconds (this is denoted as the *Optimistic unchoke interval*)

        Exactly one additional interested remote peer is unchoked at random. In other words, the optimistic unchoke slot is assigned. This optimistic unchoke has two purposes. First, to evaluate the upload capacity of new peers. Second, to provide the new peers with their first piece as soon as possible [8, 9].

Until the content download is complete.

End.

**2.2.2. Piece selection policy**

Now the focus is on the piece selection policy of BitTorrent. To begin, it is important to stress the fact that to request data pieces, a local peer has first to be unchoked by a remote peer. More precisely, a local peer is unchoked by a remote peer and, only then, may request data pieces.

The piece selection policy of BitTorrent is explained in Algorithm 2 and consists of a set of other specific policies, which map to three distinct stages [8, 9]: (i) initiating the download – the *random first policy*; (ii) normal operation – the *rarest first policy*; (iii) pulling down the last remaining pieces – the *end game mode policy*.

Nonetheless, it is valuable to add that if at least one block of a piece has been requested, then the other blocks of this same piece are requested with the highest priority. This is called the *strict priority policy* and is considered throughout the whole piece downloading operation, no matter other policy is also being deployed simultaneously [8, 9].

**Algorithm 2: Piece Selection Policy of BitTorrent**
Begin

Repeat

    *Step 1* If the peer has downloaded strictly less than four pieces, then it chooses randomly the next piece to be requested. This is called the *random first policy*. Once it has downloaded at least four pieces, it switches to the *rarest first policy*.

    The rationale behind the above step lies in the fact that rare pieces are usually only available on a single peer, so the downloading process takes longer than that for pieces which are present on multiple peers, what enables blocks to be downloaded from different peers at a time, i.e., *parallel download*. Besides, a peer without a piece cannot reciprocate. It must wait for an optimistic unchoke. In fact, the first piece is most of the time received from peers that perform optimistic unchokes [8, 9].

    *Step 2* Else if all blocks have already been requested by the local peer but have not been received, then the peer requests all blocks not yet received to all peers in its peer set that have the corresponding blocks. This is called the *end game mode policy*.

    Under this policy, when a block is received, the peer cancels its request to all peers in its peer set that have the pending request [8]. This policy is deployed when the download is to finish and hence the peer does not need to switch to any other policy. A peer never requests blocks of a same piece to more than one peer, unless it is under the *end game mode policy*.

    *Step 3* If neither Step 1 nor Step 2 takes place, then the peer selects the next piece to download at random in its rarest pieces taking into account the pieces available at the neighbour that unchokes it, i.e., it follows the behaviour of the rarest first policy. Note that, as a local peer is first unchoked by a remote peer and, only then, may request the piece, the rarest piece available on that remote peer might not be the globally rarest one.

Until the content download is complete.

End.

Also, if a peer ever happens to be choked during a piece download and this piece is not available on any other peer, the next time this peer is unchoked, a new piece will be then requested. As early as the partially downloaded piece is available again in the swarm, this peer will then request the remaining blocks with the highest priority [8, 9].

As already mentioned, a local peer sends requests to remote peers that are unchoking it. So, if $m$ remote peers unchoke the local peer, it may then potentially send requests to all these $m$ remote peers, i.e., *parallel download*. Consider, for example, the extreme case in which the local peer is the only leecher and all remote peers are seeds. Thus, at most, the local peer downloads from as many remote peers as it has in its peer set.

Still, since a remote peer sends requests to all remote peers that unchoke it and requests with highest priority blocks of partially downloaded pieces, due to the *strict priority policy*, the remote peer may then request blocks of the same piece to different remote peers. However, the remote peer never duplicates a request for the same block, but under the *end game mode policy*.

Lastly, note that a peer may request blocks of different pieces to a single remote peer or to different remote peers. Consider the case of a single remote peer: recall that the idea is to always keep $k$ block requests pipelined at once on the remote peer ($k$ is typically set to five); now suppose that there are $y$ blocks remaining to finish the download of the ongoing rarest piece and $y < k$. So there may still be $(k - y)$ block requests for the next rarest piece in order to guarantee $k$ block requests pipelined at once. Now consider the case of different remote peers: recall that the next piece to be selected by the local peer is the rarest piece available on the remote peer that has just unchoked it. Suppose that $j$ remote peers have just unchoked the local peer and that these remote peers have different rarest pieces. It follows that the remote peer will request blocks of $j$ different pieces at once.

To end this subsection, the general understanding of the piece selection policy is succinctly depicted in Figure 3.

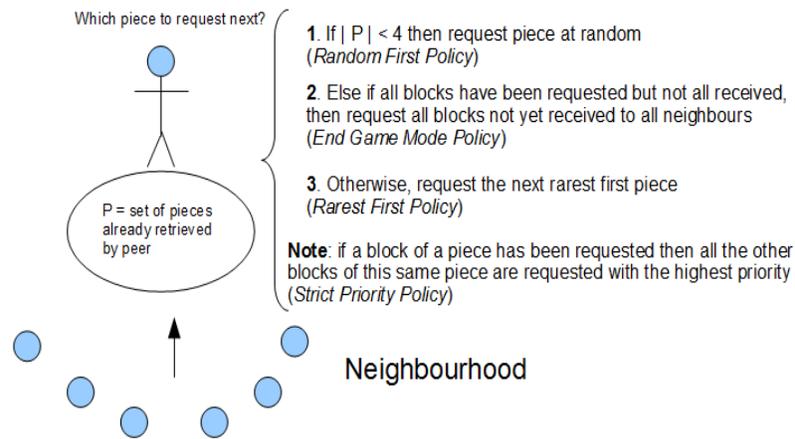

Figure 3. General understanding of the piece selection strategy.

## 2.3. Reciprocity

Reciprocity mechanisms are categorized in two kinds: direct and indirect. When a peer employs peer selection based on direct reciprocity, it unchokes other peers that have recently sent data to it at the highest rates. On the other hand, when a peer uses peer selection based on indirect reciprocity, it unchokes other peers that have sent data to others at the highest rates [1, 10].

As already mentioned, the BitTorrent' peer selection policy implements a direct reciprocity denoted as tit-for-tat strategy. Even though it provides more incentives for cooperation and may be adequate for traditional file transfer, there are two points with respect to on-demand streaming [1, 2, 6] that must be considered.

First, considering a client´s sequential data retrieving, i.e., no interactivity, peers that arrived later cannot reciprocate peers that arrived earlier, since peers that arrived later will always have a playback position behind and, hence, no data that interest those peers that arrived earlier. So, the incentives are not the same as in traditional transfer [2, 6].

Second, for on-demand streaming, there is no use for a peer to have a download rate above the file playback rate. However, under tit-for-tat, this fact is ignored and there may be a situation in which there is enough aggregate upload capacity to serve all system peers, but not all peers have an acceptable quality of service (QoS), since high-capacity peers are often favoured [2, 6].

## 3. RELATED WORK

Because of being relatively recent and have already been proven to be efficient, the proposals that follow represent some of the state-of-art schemes of the literature, concerning peer selection policies used by BitTorrent-like protocols.

Legout et al. [9] possibly examine the first modification of the original BitTorrent protocol. Their proposal considers changes exclusively when in seed state. Four remote peers may be unchoked and interested at the same time. Every interval of 10 seconds, the unchoked and interested remote peers are ordered with respect to the time they were last unchoked. The goal is to prioritize the remote peers which have been unchoked more recently. In case there is a tie, the remote peers are ordered according to the upload rate (i.e., from seed to remote peers), higher rates first. For two successive intervals of 10 seconds each, the three first peers are kept unchoked and an additional fourth peer that is choked and interested is selected at random to be unchoked. For the third period of 10 seconds, the four first peers are kept unchoked. This proposal does not focus on the time constraints of on-demand streaming applications, but mainly on the replication efficiency.

Comparing to the original BitTorrent, Shah et al. [11] propose to use the optimistic unchoke slots more frequently. More precisely, they introduce a *sliding window* which contains the next *w* pieces to be consumed by a client. They use a policy in which at the beginning of every window playback, peers perform a new optimistic unchoke. This policy therefore gives more free tries to a larger number of peers in the swarm to download pieces which they can use to share later. The point is that increasing the number of unchokes does not provide incentives for cooperation; besides, a peer may suffer from QoS degradation because of altruism.

The proposal of Mol et al. [12] basically differs from the original BitTorrent with respect to the principle of reciprocity, as explained in the following. For the three regular unchoke slots, each peer uses indirect reciprocity. As for the optimistic unchoke slots (one at a time), the assignment is similar to that of BitTorrent, but the optimistic interval is of $2\delta$ seconds ($\delta$ is typically set to 10). Exceptionally, if there is enough spare capacity, a peer is allowed to open two additional regular unchoke slots.

D'Acunto et al. [2] propose three schemes mainly devoted to heterogeneous environments, i.e., peers may have different upload and/or download capacities among themselves. They are all based on the idea of adjusting the number of upload slots a peer opens. The first scheme uses mathematical formulas which consider, e.g., the peer's upload capacity and the video playback rate. The idea of the second scheme is to have peers dynamically adjusting the number of their optimistic slots to their current QoS. Finally, the third scheme is based on the idea of giving priority to newcomers when performing optimistic unchokes. These schemes may result intricate for real implementation since they need on-line computation.

Rocha et al. [6] use the same number of regular and optimistic unchoke slots, respectively, as in BitTorrent. They though conjecture that, to efficiently select neighbours, the following values should be obtained: (1) how much the piece requests diverge from each other considering the arrival times; (2) how much of the retrieved file segments may be effectively shared; (3) how often each file position is requested. An algorithm for neighbour selection is then proposed considering these three values as main guidelines. The overall idea is to prioritize as neighbours those peers that result in neighbour sets with the lowest possible dispersion, given that they have likely retrieved most of the first file positions.

Lastly, to end this section, Table 1 brings a summary of the proposals briefly explained above as well as of the original BitTorrent. The goal is to outline the main points in common as well as the main differences among them. Within this context, it is worth adding that all these proposals, except that of D'Acunto et al. [2], employ a total of four upload slots. This mainly lies on the observation that increasing the number of peers served at the same time has the side effect of decreasing the service rate to each peer [13, 14].

Table 1. Main proposals in the literature.

| Ref. | # Reg. slots per period | # Opt. slots per period | Unchoke criterion (leecher) | Unchoke criterion (seed) | Overall key goal | Overall negative aspect |
|---|---|---|---|---|---|---|
| [8] | 3 / 10 sec | 1 / 30 sec | direct reciprocity | direct reciprocity | fast replication | does not focus on time constraints |
| [9] | 3 / 10 sec | 1 / 30 sec | direct reciprocity | time when unchoked | fast replication | does not focus on time constraints |
| [11] | 3 / 10 sec | 1 / window playback | direct reciprocity | direct reciprocity | uniform service | may be considered as too altruistic |
| [12] | 3 / 10 sec *may still use 2 extra slots | 1 / 20 sec | indirect reciprocity | indirect reciprocity | uniform service | relies on indirect information |
| [2] | may be dynamically adjusted | | mainly based on direct reciprocity | mainly based on direct reciprocity | uniform service | relies on formulas on current state |
| [6] | 3 / 10 sec | 1 / 30 sec | direct reciprocity + dispersion | Direct reciprocity + dispersion | uniform service | relies on formulas on current state |

## 4. NOVEL PROPOSALS

This section presents the three novel proposals, namely Select Balanced Neighbour Policy (SBNP), Select Optimistic Neighbour Policy (SONP), and Select Regular Neighbour Policy (SRNP). For ease of presentation and objectivity, only the steps that are modified with respect to Algorithm 1 (Subsection 2.2.1) are outlined in the algorithms to appear herein.

The SBNP proposal, in Algorithm 3, basically combines the proposals of Cohen [8] and Legout et al. [9], respectively, in addition to equally sharing the total upload bandwidth among regular and optimistic time slots, respectively. The general goal is to prioritize the remote peers which have been unchoked more recently. To eliminate the negative aspect mentioned for the proposals of Cohen [8] and Legout et al. [9], that is, no time constraints, a *sliding window* should be defined in the piece selection algorithm. This sliding window should be set to contain the data pieces that are the nearest ones to the current file playback point of the peer requesting data. When the remote peer requests a piece, it necessarily requests a piece that is within the slide window and, hence, very near to its current playback point. In this way, the on-demand time constraints are likely to be met.

> **Algorithm 3: SBNP proposal**
>
> <u>Step 2</u> The local peer's upload capacity is divided into four upload slots. Two slots are denoted as *regular unchoke slots*, and two slots are denoted as *optimistic unchoke slots*.
>
> <u>Step 3.2</u> Else if the local peer is in *seed state*: all interested remote peers are ordered according to the time they were last unchoked, most recently unchoked peers come first. The goal is to prioritize the remote peers which have been unchoked more recently. In case there is a tie, the remote peers are ordered according to the upload rate (from seed to remote peers), higher rates first.

The SONP proposal is presented in Algorithm 4. It aims to take into account the overall benefit of Shat et al. [11] and Mol et al. [12], respectively. That is, to provide more opportunities for newcomers. Nevertheless, it is more stable and conservative since the number of optimistic slots is fixed, i.e., there is no extra slot, and the time cycle for optimistic-slot assignment is set to $2\delta$ seconds, where $\delta$ is typically set to 10. This feature helps to avoid the problem of being too altruistic. Besides, this proposal solely considers direct reciprocity and, hence, does not rely on indirect information.

> **Algorithm 4: SONP proposal**
>
> <u>Step 2</u> The local peer's upload capacity is divided into four upload slots. One slot is denoted as *regular unchoke slot*, and three slots are denoted as *optimistic unchoke slots*.
>
> <u>Step 3.2</u> Else if the local peer is in *seed state*: all interested remote peers are ordered according to the time they were last unchoked, most recently unchoked peers come first. The goal is to prioritize the remote peers which have been unchoked more recently. In case there is a tie, the remote peers are ordered according to the upload rate (from seed to remote peers), higher rates first.
>
> <u>Step 4</u> Every $2\delta$ seconds, three additional interested remote peers are unchoked at random. In other words, three optimistic unchoke slots are assigned.

By its turn, the SRNP proposal comes in Algorithm 5. It mostly aims to simultaneously consider the general ideas of D'Acunto et al. [2] and Rocha et al. [6], namely to capture the dynamics of the system in order to better select the peers to which send data. Nevertheless, it may be judged as more stable and conservative since it does not deploy formulas that may result intricate to be computed in an online manner. This feature hopefully guarantees a more efficient decision-making process for the peer selection policy.

Finally, to end this section, Table 2 makes a summary of the three novel proposals. The goal is to outline their main characteristics. Besides, it is also mentioned the key design thought of each proposal as well as the references on which each of them is based.

---

**Algorithm 5: SRNP proposal**

**Step 2** The local peer´s upload capacity is divided into four upload slots. Three slots are denoted as *regular unchoke slots*, and one slot is denoted as *optimistic unchoke slot*.

**Step 3.1** If the local peer is in *leecher state*: all interested remote peers are ordered according to their upload rate to the local peer, i.e., rate at which the remote peer has sent data to the local peer. They are then ordered again now according to the time they last received pieces, so that peers that have not recently received pieces will come first. Finally, the three first peers of the list are then unchoked. In other words, the three regular unchoke slots are assigned to the fastest peers that are more time without receiving a piece.

**Step 3.2** Else if the local peer is in *seed state*: all interested remote peers are ordered according to the time they were last unchoked, most recently unchoked peers first. The goal is to prioritize the remote peers which have been unchoked more recently. In case there is a tie, the remote peers are ordered according to the upload rate (from seed to remote peers), higher rates first.

---

## 5. PERFORMANCE EVALUATION

### 5.1. Scenario characterization

Many works in the literature are especially dedicated to the task of characterizing BitTorrent systems. To name a few, there are the works of [15-19]. In spite of being deeply intricate, this task is indeed invaluable. It mainly enables researchers to adequately choose system characterization parameters and set their numerical values to carry out simulation studies and experiments based on real-world scenarios.

International Journal of Computer Networks & Communications (IJCNC) Vol.6, No.1, January 2014

Table 2. Novel proposals.

| Proposal | # Reg. slots per period | # Opt. slots per period | Unchoked riterion (leecher) | Unchoked riterion (seed) | Key thought | Basis (references) |
|---|---|---|---|---|---|---|
| SBNP | 2 / 10 sec | 2 / 30 sec | upload rate | time being unchoked | number of slots are equal | [8] and [9] |
| SONP | 1 / 10 sec | 3 / 20 sec | upload rate | time being unchoked | prioritize optimistic slots | [11] and [12] |
| SRNP | 3 / 10 sec | 1 / 30 sec | upload rate + time receiving piece | time being unchoked | capture system dynamics | [2] and [6] |

For example, Souza e Silva et al. [15] developed a crawler to monitor one of the most popular Torrent Search Engines (Torlock.com) to determine the daily size of all swarms announced in the website (around 150,000 swarms) for ten consecutive days. Their final results indicate that most of the swarms are very small: about 73% of the total is formed by less than 10 peers, and 58% have less than five peers. This fact implies that, to analyze ordinary swarms, a size of less than 10 peers should be then considered.

On the other hand, the large measurement carried out in the work of Hoßfeld et al. [7] indicates that there is a clear relationship between the content type and the swarm size, as exemplified in the following. For TV series, the average size is 15.53 peers per swarm and, for Movies and Music files, the averages are 25.46 and 9.76 peers, respectively, per swarm. These figures suggest that non-video contents tend to be less popular than video contents. So, in case there is more interest on the replication of video contents, larger swarms should be then the focus.

Another important characterization parameter refers to the content size. For example, by means of measurement and analysis of real BitTorrent swarm systems, Wang et al. [19] have concluded that 90% of video contents are larger than 100 MB, there are 5% of the video contents with size being larger than 10 GB, and the maximum video reaches nearly 20 GB. On the other hand, they have also found that only 30% of the non-video contents are larger than 100 MB, and over 50% of non-video contents are less than 20 MB, whereas those small contents are very few in the existing video file swarms.

Being aware of the above, Table 3 is presented. Its numerical values are set later in Subsection 5.3 and are chosen with two goals in mind. First, to represent real typical BitTorrent swarm scenarios. Second, to allow simulation studies focused on the objective of this work: to design peer selection policies for BitTorrent-like protocols devoted to interactive on-demand streaming systems.

214

International Journal of Computer Networks & Communications (IJCNC) Vol.6, No.1, January 2014

Table 3. Key characterization parameters.

| Symbol | Definition |
|---|---|
| $m$ | Number of leechers in the swarm |
| $n$ | Number of seeds in the swarm |
| $O_s$ | Content size, measured in bytes. |
| $p_s$ | Piece size, measured in bytes. |
| $b_s$ | Block size, measured in bytes. |
| $R_{down}$ | Peer´s (leecher or seed) download rate, measured in bytes per seconds. |
| $R_{up}$ | Peer´s (leecher or seed) upload rate, measured in bytes per seconds. |

## 5.2. Simulation setup and performance metrics

The three novel proposals are modeled on top of the simulation tool Tangram-II [20]. It combines a sophisticated user interface based on an object-oriented paradigm and solution techniques for performance and availability analysis. The user specifies a model in terms of objects that interact via a message exchange mechanism. Once the model is compiled, it can be either solved analytically, if it is Markovian or belongs to a class of non-Markovian models, or solved via simulation. There are several solvers available to the user, both for steady state and transient analysis, respectively.

The modeling herein solely refers to the peer selection policy, i.e., it does not take into account the piece selection policy. Therefore, the results, analysis and conclusions are focused on system optimization provided exclusively by the peer selection policy. Additionally, the system to be evaluated in the simulations is considered to be in steady state, i.e., although peers might join and leave the system, the total number of peers remains constant. Lastly, the simulation results have 95% confidence intervals that are within 5% of the reported values.

The following performance metrics are used in the simulations: *service time*, *number of clients served* and *efficiency retrieving coefficient* (ERC). The first refers to the period of time the client has to stay in the system to receive the wanted data. The second refers to the total number of clients that are served during the entire simulation time. The third metric is calculated by means of Equation (1) and says how efficient a peer´s data retrieving process is compared to that employing an exclusive data-delivery channel. Recall that a swarm peer eventually needs to take turns with other system peers in order to download the wanted content, i.e., the content downloading process may be interrupted several times before it finally ends.

$$ERC = \frac{\sum_{k=1}^{m} D_k}{\sum_{k=1}^{m} T_k}, \qquad (1)$$

where: $m$ is the number of leechers in the swarm; $D_k$ is the data retrieving time of leecher $k$, given that it has an exclusive data-delivery channel, i.e., it does not suffer from any interruptions. The value of $D_k$ is computed by the ratio $O_s/R_{down}$; $T_k$ is the data retrieving time of leecher $k$, given that it participates in the swarm and, hence, suffers from interruptions. $T_k$ is obtained from the simulations. Lastly, note that the value of ERC lies into the interval [0, 1], and the closer ERC is to 1, the more efficient the proposal being evaluated is.

Lastly, to obtain the most possible accurate results for the performance metrics herein defined, the peer and piece selection policies, respectively, should be both modeled. This is because it is very likely to exist inter-influences between them, i.e., the piece selection influences the peer selection and vice-versa. Nevertheless, these possible inter-influences are mostly quantitative. This implies that they may modify the absolute values but not the qualitative and relative values. Hence, the results herein obtained are indeed sufficient and satisfactory for a truly impartial and competitive analysis. To model the piece selection policy is a complementary work deliberately left for the future.

## 5.3. Results and analysis

Results and analysis are separated into three subsections, as explained in the following. Subsection 5.3.1 is dedicated to an overall competitive analysis among the three proposed policies. To this end, different file replication scenarios, mainly defined in function of the parameters of Table 1, are examined. Subsection 5.3.2 is specifically devoted to evaluating the scalability, in terms of content size, of the most efficient proposal resulting from the analysis of Subsection 5.3.1. To this end, replication scenarios owning contents of different sizes are considered. Finally, Subsection 5.3.3 takes into account the client´s interactive behaviour and, again, solely the most efficient proposal coming out from Subsection 5.3.1. The goal is to quantify the influence of the interactive behaviour on the proposal´s performance.

### 5.3.1. Overall competitiveness

Table 4 presents typical values of the parameters defined in Table 3. They are based on works of the literature such as [3, 4, 7, 15-18]. As it can be seen, there are four scenarios, representing swarms, categorized in accordance with the content type: Music Files (MF); TV Series (TS); Movies (M); All Media (AM). The last one refers to swarms that are not restricted to any type of content.

Figures 4(a) and 4(b) bring the values of ERC and the total number of clients served, respectively, considering the three novel proposals deployed at the four distinct scenarios. Note that a similar number of clients served suffices to assure a truly fair ERC-based comparison among the proposals under a same scenario.

The results observed for the metric ERC clearly indicate that there is no noticeable difference in performance among the proposals. Given that four uploading slots are used in total, the rationale behind this fact is that neither the quantity of slots assigned (either regular or optimistic) for each peer swarm nor the time cycle assigned for slot re-evaluation are that significant to impact on the final performance.

In other words, considering four data slots for each swarm peer under typical scenarios of file replication, the corresponding slot categorization, either in regular or optimistic, does not really affect the proposal´s overall performance. Nevertheless, in case of choosing one of these proposals, the SBNP may be pointed out as the most adequate alternative, since it is the one that best preserves the natural equilibrium between direct and indirect reciprocities, thus equally sharing the peer´s upload bandwidth among faster and slower peers.

### 5.3.2. Scalability

Four scenarios, categorized in function of the content size, are herein examined. These scenarios are: Ordinary – $O_s$ = 20.0 MB; Medium – $O_s$ = 200.0 MB, Large – $O_s$ = 2000.0 MB; and Very Large – $O_s$ = 20000.0 MB. Note that the content size varies in at least one-order of magnitude from one scenario to another. The goal is to see whether this variation may influence on the SBNP´s performance, identified as the best proposal in the last subsection. The other characterization parameters are the same of Table 4 and are set as in the AM scenario.

216

From Figure 5(a), the results of the metric ERC explicitly indicate that there is no noticeable difference in performance. This means that the SBNP proposal is indeed a very scalable solution for content replication, independently of the content size itself. This is a very important finding and may not be neglected at all when deciding about real solution implementations.

Table 4. Key characterization parameters for performance evaluation.

| Parameters | Replication scenarios | | | |
| --- | --- | --- | --- | --- |
| | Music files (MF) | TV series (TS) | Movies (M) | All media (AM) |
| $m$ | 10 | 15 | 25 | 7 |
| $n$ | 1 | 1 | 1 | 1 |
| $O_s$ | 10.0 MB | 100.0 MB | 200.0 MB | 20.0 MB |
| $p_s$ | 256 B | 256 B | 256 B | 256 B |
| $b_s$ | 16 B | 16 B | 16 B | 16 B |
| $R_{down}$ | $10^4$ B/sec | $10^5$ B/s | $10^4$ B/s | $10^4$ B/s |
| $R_{up}$ | | | | |

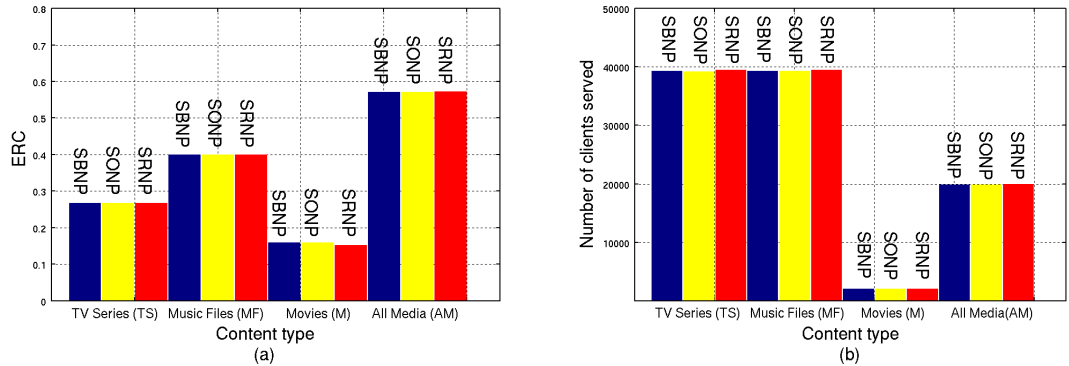

Figure 4. (a) ERC in distinct scenarios; (b) Total number of clients served in distinct scenarios.

Nevertheless, it is worth stressing that a same value of ERC in distinct scenarios does not necessarily mean that the client´s service time is identical in all of the distinct scenarios. The service time is a function of the client´s download rate. Since, in the simulations, all scenarios are considered to have a same client´s download rate, the service time tends to be the shortest for the Ordinary scenario and the longest for the Very Large scenario. This fact may be observed in Figure 5(b). Clearly, the service time varies at about one-order magnitude from one scenario to another, considering an ascending-size order.

### 5.3.3. Interactivity

It has already been observed that the more interactive the client is, the less data he requests and, hence, less time he is likely to stay in the system. Also, the more interactive the client is, the more playout interruptions he tends to suffer during his system sessions [6, 21, 22].

The playout interruptions may result, for instance, from two different situations. First, the more interactive the client is, the more distinct non-sequential file playout points he is likely to need. This means there have to be waiting-time periods to first search for the non-sequential data before it may be really retrieved. Second, the client may deliberately decide to simply pause the session and resume to it after some time. From this, it comes that the playout interruptions directly impact on the client´s *slot occupancy*, which computes how much of the slot itself, in time percentage, is effectively used for real data retrieving.

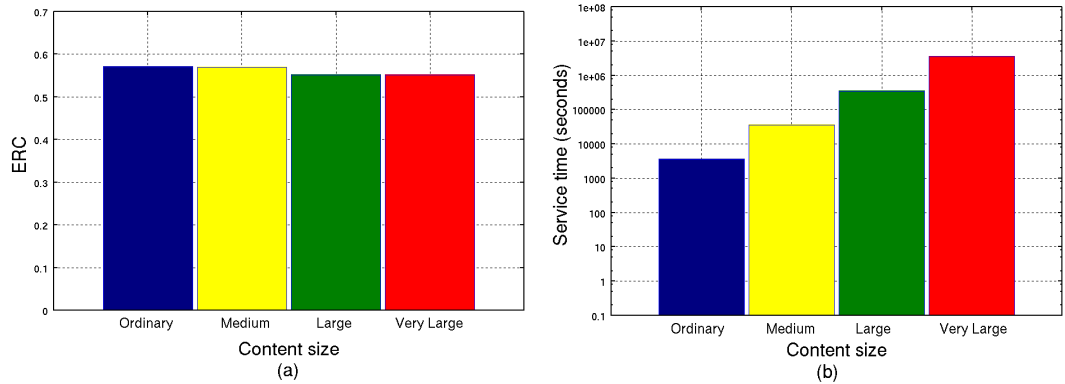

Figure 5. (a) ERC for distinct content sizes; (b) Services times for distinct content sizes.

Considering the above, the following categorization for the client´s interactivity profile is admitted: High Interactivity (HI) – the slot occupancy is 40%; Medium Interactivity (MI) – the slot occupancy is 60%; Low Interactivity (LI) – the slot occupancy is 90%; No Interactivity (NI) – the slot occupancy is 100%. These numerical values are chosen to represent real-world swarms and are based on works of the literature such as [6, 21, 22].

Four scenarios, categorized in function of the client´s interactivity profile, are then examined in this section. Each scenario refers to one of the profiles. Besides, it is considered that the clients retrieve the entire file. The goal is to see how much the client´s interactivity may influence on the SBNP´s performance, identified as the best proposal in Subsection 5.3.1. The other parameters are the same of Table 4 and are set in accordance with the AM scenario.

From Figure 6(a), the final values of ERC indicate that the more interactive the client is, the less efficient the proposal becomes. By its turn, Figure 6(b) says that there are longer service times for the more interactive clients. The rationale behind these two general observations is that: the more interactive client needs more time slots than a less interactive client in order to retrieve the same quantity of data. Recall that all clients want to retrieve the same quantity of data.

It is worth mentioning that the final values of ERC present a sort of linear dependence on the slot occupancy. For instance, when the slot occupancy is reduced from 100% to 40%, there is a corresponding reduction of about 60% at the final value of ERC. In the case of the service time, there is an inverse dependence that follows a non-linear pattern.

Still, deploying a local buffer on the client´s side to save all data retrieved by him may potentially optimize the values obtained for the metrics ERC and service time, respectively. This is because it is very likely that the interactive behaviour results in data requests for the same playout points during the client´s session. For instance, when seeing a movie, a client may decide to review a specific scene he likes most; in this case, if the data is saved in a local buffer, the data request may be promptly answered since there is no network access because the data is already locally available. This implies a zero waiting-time service, what directly impacts on the values of ERC and service time. This analysis is deliberately left for future work.

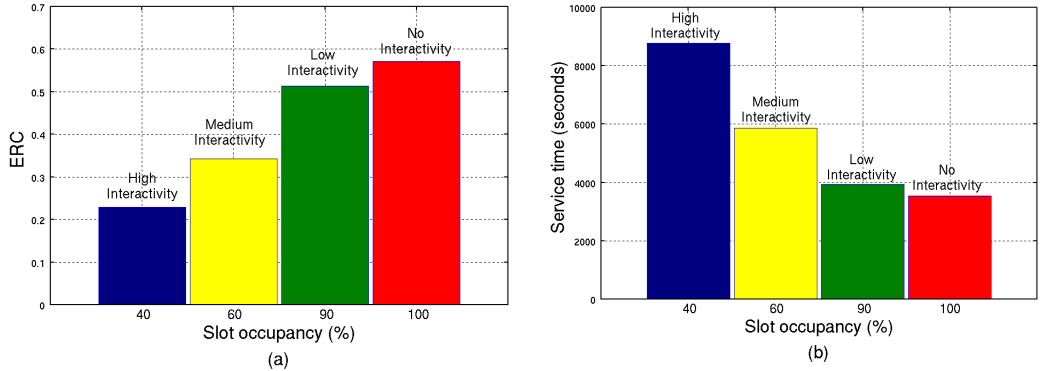

Figure 6. (a) ERC for interactivity profiles; (b) Service time for interactivity profiles.

## 6. CONCLUSIONS AND FUTURE WORK

This article introduced three novel peer selection policies for the design of BitTorrent-like protocols, namely Select Balanced Neighbour Policy (SBNP), Select Regular Neighbour Policy (SRNP), and Select Optimistic Neighbour Policy (SONP). They are especially devoted to Internet on-demand streaming systems, considering the client´s interactive behaviour, and were validated by means of simulations on a variety of file replication scenarios. *Service time*, *number of clients served* and *efficiency retrieving coefficient* (ERC) were the performance metrics computed in the simulations.

In general, the novel proposals showed to be efficient solutions, independently of the size and type of content to be replicated. For instance, the ERC reached values up to 0.571, indicating a satisfactory data retrieving time, and the number of clients served reached up to 39414, mainly evidencing scalability and robustness. Besides, the following more specific results may be outlined: (i) given four data slots for each system peer and considering typical scenarios of file replication, the corresponding slot categorization, either in regular or optimistic, does not affect the proposals´ performance to a real significant extent; (ii) to preserve the natural equilibrium between direct reciprocity and indirect reciprocity, the SBNP proposal may be considered as the most adequate proposal. This is a very important finding and may not be neglected at all when deciding about a real solution implementation; lastly, (iii) under interactive scenarios, deploying local buffers on the client´s side may be an effective procedure to prevent any performance degradation.

Future work in this research field may specially include: (i) to implement the proposals herein presented together with proven-efficient piece selection policies of the literature, resulting in novel BitTorrent-like proposals for on-demand streaming systems. These proposals should be then compared with others of the literature; (ii) to quantify the data-retrieving efficiency that may be granted to BitTorrent-like protocols by the eventual deployment of local buffers on the client´s side, under interactive scenarios; (iii) to look into the *peer request problem* [23], the *service request problem* [23], and the traffic locality of P2P file swarming systems [19, 24] to see whether there is still space to optimize the BitTorrent's peer and piece selection policies [25], respectively, targeting at on-demand streaming solutions; (iv) lastly, to examine the influences of the Network Address Translation (NAT) service [26] as well as of traffic hierarchical scheduling [27] on the design of P2P on-demand streaming protocols.

## ACKNOWLEDGEMENTS

We are in debt to Professor Arnaud Legout from the INRIA Sophia Antipolis, France, who helped us to clarify important features of the BitTorrent protocol´s core algorithms. Also, we are in debt to Doctor Guilherme Dutra Gonzaga Jaime from the Nuclear Engineering Institute, Brazil, for patiently answering our questions about the simulation tool Tangram.

## REFERENCES

[1]     D'Acunto, L., Chiluka, N., Vinkó, T. & Sips, H. (2013) "BitTorrent-like P2P approaches for VoD: A comparative study", *Computer Networks*, Vol. 57, No. 5, pp 1253 – 1276.

[2]     D'Acunto, L., Andrade, J. & Sips, H. (2010) "Peer selection strategies for improved QoS in heterogeneous BitTorrent-like VoD systems", *IEEE International Symposium on Multimedia*, Taichung, Taiwan.

[3]     Xia, R. L. & Muppala, J. K. (2010) "A survey of BitTorrent performance", *IEEE Communications surveys & tutorials*, Vol. 12, No. 2, pp 140 – 158.

[4]     Guo, L., Chen, S., Xiao, Z., Tan, E., Ding, X. & Zhang, X. (2007) "A performance study of BitTorrent-like peer-to-peer systems", *IEEE JSAC*, Vol. 25, No. 1, pp 155 – 169.

[5]     Hoffmann, L. J., Rodrigues, C.K.S. & Leão, R. M. M. (2011) "BitTorrent-like protocols for interactive access to VoD systems", *European Journal of Scientific Research*, Vol. 58, No. 4, pp 550 – 569.

[6]     Rocha, M. V. M. & Rodrigues, C. K. S. (2013) "On Client's interactive behaviour to design peer selection policies for BitTorrent-like protocols", *International Journal of Computer Networks & Communications (IJCNC)*, Vol. 5, No. 5, pp 141 – 159.

[7]     Hoßfeld, T., Lehrieder, F., Hock, D., Oechsner, S., Despotovic, Z., Kellerer, W. & Michel, M. (2011) "Characterization of BitTorrent swarms and their distribution in the Internet", *Computer Networks*, Vol. 55, No. 5, pp 1197 – 1215.

[8]     Cohen, B. (2003) "Incentives build robustness in BitTorrent", *First Workshop on Economics of Peer-to-Peer Systems*, Berkeley, EUA.

[9]     Legout, A., Urvoy-Keller, G. & Michiardi, P. (2006) "Rarest first and choke algorithms are enough", *6th ACM SIGCOM Conference on Internet Measurement*, Rio de Janeiro, Brazil.

[10]    Menasché, D., Massoulié, L. & Towsley, D. (2010) "Reciprocity and barter in peer-to-peer systems", *29th Conference on Information communications*, San Diego, CA, USA.

[11]    Shah, P. & Pâris, J.-F. (2007) "Peer-to-Peer multimedia streaming using BitTorrent", *IEEE International Performance, Computing, and Communications Conference (IPCCC)*, New Orleans, EUA.

[12]    Mol, J., Pouwelse, J., Meulpolder, M., Epema, D. & Sips, H. (2008) "Give-to-Get: Free-riding-resilient video-on-demand in P2P systems", *SPIE MMCN*, San Jose, California, USA.

[13]    Yang, X. & de Veciana, G. (2004) "Service capacity of peer-to-peer network", *23rd Annual Joint Conference of IEEE Computer and Communications Societies (INFOCOM 2004)*, Vol. 4, pp 2242 – 2252.

[14]    Biersack, E., Rodriguez, P. & Felber, P. (2004) "Performance analysis of peer-to-peer networks for file distribution", *Lectures Notes in Computer Science (LNCS)*, Vol. 3266, pp 1 – 10.

[15]    de Souza e Silva, E., Leão, R., Menasché, D. & Rocha, A. (2013) "On the interplay between content popularity and performance in P2P systems", *Lecture Notes in Computer Science (LNCS)*, Vol. 8054, pp 3 – 21.

[16]    Murai, F., Rocha, A., Figueiredo, D. & Souza e Silva, E. (2011) "Heterogeneous download times in a homogeneous BitTorrent swarm", *Computer Networks*, Vol. 56, No. 7, pp 1983 – 2000.

[17]    Ye, L., Zhang, H., Li, F. & Su, M. (2010) "A measurement study on BitTorrent system", *International Journal of Communications, Network and System Sciences*, Vol. 3, pp 916 – 924.

[18]    Varvello, M., Steiner, M. & Laevens, K. (2012) "Understanding BitTorrent: a reality check from the ISP's perspective", *Computer Networks*, Vol.56, No. 40, pp 1054 – 1065.

[19]    Wang, H., Liu, J. & Xu, K. (2012) "Understand traffic locality of peer-to-peer video file swarming", *Computer Communications*, Vol. 35, No. 15, pp 1930 – 1937.

[20] De Souza e Silva, E., Figueiredo, D. & Leão, R. (2009) "The Tangram-II Integrated Modelling Environment for Computer Systems and Networks", *ACM SIGMETRICS Performance Evaluation Review*, Vol. 36, No. 4, pp 45 – 65.
[21] Costa, C., Cunha, I., Borges, A., Ramos, C., Rocha, M., Almeida, J., & Ribeiro-Neto, B. (2004) "Analyzing Client Interactive Behavior on Streaming Media Servers", *13th WWW Conference*, New York, USA.
[22] Hoffmann, L. J., Rodrigues, C.K.S. & Leão, R. M. M. (2008) "On Scalable Interactive Video-On-Demand Services", *European Journal of Scientific Research*, Vol. 21, No. 4, pp 662 – 686.
[23] Yang, Y., Chow, A., Golubchik, L. & Bragg, D. (2010) "Improving QoS in bittorrent-like VoD systems", *Proceedings of the IEEE INFOCOM*, San Diego, CA, USA.
[24] Le Blond, S., Legout, A. & Dabbous, W. (2011) "Pushing BitTorrent locality to the limit", *Computer Networks*, Vol. 55, No. 3, pp 541 – 557.
[25] Atlidakis, V., Roussopoulos, M. & Delis, A. (2014) "EnhancedBit: Unleashing the potential of the unchoking policy in the BitTorrent protocol", *Journal of Parallel and Distributed Computing*, Vol. 74, No. 1, pp 1959 – 1970.
[26] Masoud, M. Z. M (2013) "Analytical modelling of localized P2P streaming systems under NAT consideration", *International Journal of Computer Networks & Communications (IJCNC)*, Vol. 5, No. 3, pp 73 – 89.
[27] Barhoun, R., Namir, A. & Barhoun, A. (2013) "Analysis of hierarchical scheduling for heterogeneous traffic over network", *International Journal of Computer Networks & Communications (IJCNC)*, Vol. 5, No. 3, pp 103 – 116.

**Author**

Carlo Kleber da S. Rodrigues received the B.Sc. degree in Electrical Engineering from the Federal University of Paraiba in 1993, the M.Sc. degree in Systems and Computation from the Military Institute of Engineering (IME) in 2000, and the D.Sc. degree in System Engineering and Computation from the Federal University of Rio de Janeiro (UFRJ) in 2006. Currently he is Military Assessor of the Brazilian Army in Ecuador, Professor at the Armed Forces University (ESPE) in Ecuador, and Professor at the University Center 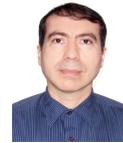 UniCEUB in Brazil. His research interests include the areas of computer networks, performance evaluation, and multimedia streaming.